\documentclass[aip,preprint]{revtex4-1}
\usepackage{graphicx}
\usepackage[space]{grffile}
\usepackage{dcolumn}
\usepackage{bm}
\usepackage{subfigure}
\usepackage{mathrsfs}
\usepackage{booktabs}
\usepackage{notoccite}
\usepackage{float}
\usepackage{placeins}
\usepackage{enumerate}
\usepackage{gensymb}
\usepackage{amsmath}
\usepackage{color}
\usepackage{siunitx}
\usepackage{mathrsfs}

\newcommand{\rom}[1]{\uppercase\expandafter{\romannumeral #1\relax}}

\begin{document}

\title{Electrochemical stability and light-harvesting ability of silicon photoelectrodes in aqueous environments}

\author{Quinn Campbell}
\email{quinn.campbell@psu.edu}
\author{Ismaila Dabo}
\affiliation{Department of Materials Science and Engineering, 
Materials Research Institute, and Penn State Institutes of Energy 
and the Environment, The Pennsylvania State University, University 
Park, PA 16802, USA}
\date{\today}

\begin{abstract}
We consider the factors that affect the photoactivity of silicon electrodes for the water-splitting reaction using a self-consistent continuum solvation (SCCS) model of the solid-liquid interface.  This model allows us to calculate the charge-voltage response, Schottky barriers, and surface stability of different terminations while accounting for the interactions between the charge-pinning centers at the surface and the depletion region of the semiconductor. We predict that the most stable oxidized surface does not have a favorable Schottky barrier, which further explains the low solar-to-hydrogen performance of passivated silicon electrodes.
\end{abstract}

\pacs{}

\maketitle

\section{Introduction}

 Hydrogen is a sustainable energy carrier that can be produced by splitting water at the surface of a photocatalytic semiconductor \cite{USDepartmentofEnergy2013,Lewis2007,Fujishima1972,Jafari2016}. Yet, most of the semiconductors that are presently well developed for photovoltaics transfer poorly to photocatalysis \cite{Leempoei1982,Chen2011,Bolts1979}. In the case of silicon, for instance, multiple factors contribute to the low solar-to-hydrogen efficiency. Notably, the electrochemical corrosion and surface restructuring of silicon electrodes are primary limitations to their use as photocatalysts. In fact, silicon is prone to oxidize in water, forming a passivating silica deposit at the solid-liquid interface \cite{Runyan2013,Graf1989}. 

While this process is known, the mechanisms that limit the photocatalytic efficiency have not been fully explored at the molecular level. The oxidation alter several properties of the interface, including the band gap, the surface states, and the Schottky barrier height. The latter is the potential difference between the bulk of the semiconductor and the surface; it provides the motive force for separating the photogenerated electrons and holes and transporting the excited charges from the electrode to the electrolyte \cite{Bisquert2014,Rajeshwar2007,Brillson2016}. Therefore, the Schottky barrier is a central indicator of the light-harvesting ability of semiconductor electrodes.

First-principles methods, with their ability to probe atomic length scales, provide an ideal approach for examining the mechanisms that underlie the activity of photoelectrodes as a function of the surface termination. While several electronic-structure methods have been developed to understand neutral photocatalytic interfaces\cite{Kharche2014,Cheng2012,Ping2015,Wu2011}, until recently, no work has included the effects of applying a potential and controlling the hydrogen activity. Here, we exploit our newly developed semiconductor-continuum methodology\cite{Campbell2017} to understand the charge-voltage response of different structures of the silicon-water interface. We extend this methodology to predict the stability and Schottky barriers of the various  terminations, showing that while the most oxidized surface is typically the most stable, it also exhibits a low Schottky barrier, which helps understand the limited efficiency of silicon for photocatalytic hydrogen generation. Our work illustrates the capabilities of first-principles methods to identify the molecular factors controlling the photoactivity of semiconductor electrodes.

\section{Background}

\label{sec:theory}

\begin{figure}
	\includegraphics[width=0.5\columnwidth]{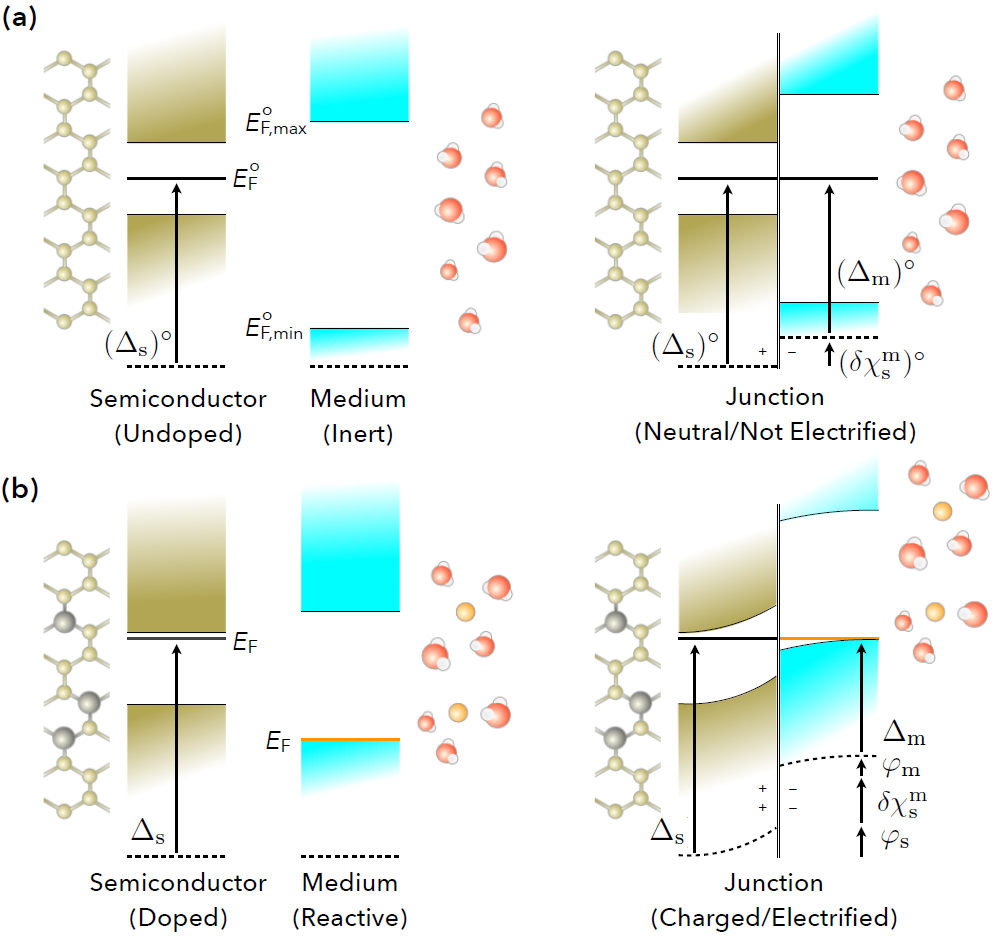}
	\caption{(a) A pristine semiconductor in contact with a 
		chemically inert medium (here, a solvent of wide redox stability window) 
		has all of its potential drop located at the interface, leading to the 
		formation of the surface dipole $(\delta\chi_{\rm s}^{\rm m})^\circ$.  
		(b) When the Fermi level of an extrinsic semiconductor and that of a 
		reactive medium (here, the same solvent with electron-accepting ionic 
		species) are initially shifted from one another, the potential drop to realign 
		the energy levels is distributed between the surface dipole 
		$\delta\chi_{\rm s}^{\rm m}$, the Schottky barrier 
		$\varphi_{\rm s}$ within the semiconductor, and the potential shift 
		$\varphi_{\rm m}$ within the medium. 
		\label{fig:realignment}}
\end{figure}

To predict the influence of surface termination on charge separation, we first examine the microscopic mechanisms that lead to the formation of the Schottky barrier. We start by focusing on the interface between an intrinsic semiconductor and a chemically inert medium; in specific terms, we consider a semiconducting electrode in contact with an ideal electrolyte, i.e., one that does not interact chemically with the electrode and is stable over a wide range of applied voltage. Since the electrolyte is, in this ideal case, insensitive to the applied voltage, the equilibration of the system will take place without constraint on the Fermi energy. Consequently, no electronic charge will be injected or withdrawn from the semiconductor. Figure \ref{fig:realignment}(a) depicts the resulting equilibrium state, where a surface dipole forms due to reorientation of the solvent molecules. This surface dipole can be expressed as $(\delta \chi_{\rm s}^{\rm m})^\circ = (\Delta_{\rm s})^\circ - (\Delta_{\rm m})^\circ$, where $(\Delta_{\rm s})^\circ$ and $(\Delta_{\rm m})^\circ$ are the differences between the Fermi levels and average electrostatic potentials for the semiconductor and the surrounding medium \footnote{In this electronic-structure study, we describe the electronic properties of the junction in terms of differences between the Fermi level and the average electrostatic potential of the two terminals, instead of electronegativity differences as is common in the experimental literature. These conventions are equivalent under an appropriate shift of the relative energy scales, as shown in Sec.~S1 of the supporting information.}.

Then, if defects are introduced in the semiconductor and chemically active ionic species are added to the embedding electrolyte, the chemical window of the solution will be reduced, causing the Fermi level of the electrode to be pinned by the chemical potential of the reaction that limits the stability of the reactive medium. This constraint leads to a different equilibrium state [Fig.~\ref{fig:realignment}(b)], where defect charge builds up within the depletion layer of the electrode and compensating ions accumulate within the double layer of the electrolyte. As a result, a large drop in the electrostatic potential is observed, corresponding to an increased surface dipole $\delta \chi_{\rm s}^{\rm m}$, a potential drop within the depletion layer of the electrode, namely, the Schottky barrier $\varphi_{\rm s}$, and another electrostatic shift in the electrical double layer $\varphi_{\rm m}$ \cite{Bockris2000a,Schmikler2010}. In this equilibrated state, the Schottky barrier $\varphi_{\rm s}$ is simply related to the surface dipole through 
\begin{equation}
\label{eqn:schottky}
\varphi_{\rm s} =\Delta_{\rm s} - \Delta_{\rm m} -\delta \chi_{\rm s}^{\rm m} - \varphi_{\rm m},
\end{equation}
where $\Delta_{\rm s}$ and $\Delta_{\rm m}$ are the differences between the Fermi level (divided by the electron charge $e_0$) and the bulk potential of the doped semiconductor and the chemically active medium, respectively. An electrostatic shift $\varphi_{\rm m}$ will also take place in the solvating medium. Since the ionic concentration of this medium is typically orders of magnitude greater than the carrier concentration of the semiconductor, the potential drop in the medium will be negligible: $\varphi_{\rm m} \ll \varphi_{\rm s}$. (In practical experiments,  $\varphi_{\rm m}$ can even be eliminated by varying the potential and pH of the solution to reach the ``zeta potential'' at which there is no interfacial charge and no accumulation of H$^+$ and OH$^-$ species \cite{Kharche2014,Cheng2012}.) 

Predicting the Schottky barrier of a semiconductor interface from first principles is challenging; since a barrier height of 0.5 V across a semiconductor with a dielectric constant of $\sim$10 and a carrier density of 10$^{16}$ cm$^{-3}$ would extend as far as $\sim$250 nm into the semiconductor, the direct quantum-mechanical modeling of Schottky barriers is prohibitively demanding. Hence, a common practice in the first-principles density-functional theory literature has been to compute Schottky barriers at electrically neutral surfaces with a simulation range on the order of $\sim$10 nm. In this approach, the neutral-interface dipole $(\delta \chi_{\rm s}^{\rm m})^\circ $ between an intrinsic semiconducting surface and an electrically neutral embedding medium is found and any remaining difference between the Fermi levels is equated to the Schottky barrier \cite{Jiao2015,Ping2015,Farmanbar2015,Goniakowskt2004,Stengel2011a,Khomyakov2009,Dong2008,McKee2003,Padilha2015,Picozzi1998,Jeon2006}:
\begin{equation}\\
\varphi_{\rm s} \approx \Delta_{\rm s}  - \Delta_{\rm m}  -(\delta \chi_{\rm s}^{\rm m})^\circ.
\end{equation}

This commonly used approach overestimates the barrier height, as can be seen by rearranging Eq.~\ref{eqn:schottky} into
\begin{equation}\\
\label{eqn:schottky_long}
\varphi_{\rm s} = [\Delta_{\rm s}  - \Delta_{\rm m}  -(\delta \chi_{\rm s}^{\rm m})^\circ ]  - [\delta \chi_{\rm s}^{\rm m} - (\delta \chi_{\rm s}^{\rm m})^\circ ].
\end{equation}
In many cases, calculating only the first half of the right hand side of Eq.~\ref{eqn:schottky_long}, as done in this conventional approach, is a reasonable approximation; the overestimation of the Schottky barrier height may be compensated to some degree by the underestimated band gap within local and semilocal density-functional theory. Yet, this approach is applicable only when $\delta \chi_{\rm s}^{\rm m} \approx (\delta \chi_{\rm s}^{\rm m})^\circ$, i.e., when the surface dipole of the charged interface under bias is close to that of the neutral interface, which is only true for interfaces that remain moderately polarized under bias. Therefore, there is a need for a computational model that  would incorporate the effects of surface termination and external bias and would predict Schottky barrier heights accurately.

To fill this gap, we apply and further develop a first-principles approach that integrates an implicit semiclassical description of the bulk semiconductor with an explicit electronic-structure treatment of the semiconductor surface. This method, detailed in the next section, allows us to  determine the potential-dependent interfacial dipole and the resulting equilibrium Schottky barriers.

\section{Model}

\subsection{Quantum--continuum embedding}

\label{sec:methods}

The proposed computational method consists of embedding an explicit quantum-mechanical description of the semiconductor surface layers into an implicit continuum model of the semiconducting and electrolytic environments. As described in Ref.~\onlinecite{Campbell2017}, this procedure exploits the self-consistent continuum solvation (SCCS) approach \cite{Andreussi2012}, which introduces dielectric cavities around each facet of the slab. The dielectric permittivity is expressed as
\begin{equation}
\epsilon (\bm{r}) =\exp[(\zeta(\bm{r}) -\sin(2\pi\zeta(\bm{r}))/2\pi)\ln\epsilon_{\rm m}]
\end{equation}
where $\epsilon_{\rm m}$ is the dielectric constant of the electrolytic medium and $\zeta(\bm{r}) =(\ln \rho_{\rm max} - \ln\rho(\bm{r}))/(\ln \rho_{\rm max} -\ln \rho_{\rm min}) $ is used as a smooth switching function, marking the transition between the quantum-mechanical and semiclassical continuum regions. Here, $\rho_{\rm min}$ and $\rho_{\rm max}$ serve as the density thresholds specifying the inner and outer isocontours of the dielectric cavity. The SCCS model also includes contributions from the external pressure, solvent surface tension, and solvent dispersion and repulsion effects. The surface tension is described by $G_{\rm cav}=\gamma S$ and the dispersion and repulsion effects by $G_{\rm dis+rep} = \alpha S + \beta V $. Here, $\gamma$ is the solvent surface tension, taken from experiment, $\alpha$ and $\beta$ are fitted parameters, and $S$ and $V$ are the quantum surface and volume of the solute, defined as $S = \int d\mathbf{r}(d\Theta/d\rho)  |\nabla \rho|$ and $V=\int d\mathbf{r} \Theta(\rho) $, where $\Theta$ is another smooth switching function, defined by $\Theta(\rho) = (\epsilon_{\rm s} - \epsilon(\rho))/(\epsilon_{\rm s}-1)$. We utilize the parameterization of Andreussi {\it et al.}, where $\rho_{\rm max}= 5 \times 10^{-3}$ a.u., $\rho_{\rm min}= 1 \times 10^{-4}$ a.u., $\gamma = 72.0$ dyn/cm, $\alpha = -22$ dyn/cm, and $\beta = -0.35$ GPa \cite{Andreussi2012}. 

A recent study has shown that the volume contribution $\beta V$ to the total energy of the interface is physically inappropriate for a slab model, as it leads to a spurious dependence of this energy on the thickness size of the slab \cite{Fisicaro2017,Andreussi2018}. However, because the simulated slabs are of comparable volume, the spurious term cancels out and does not affect the relative energies. 

We also note that the parameterization of $\rho_{\rm max}$ and $\rho_{\rm min}$ was initially developed for neutral molecular systems in water. Adopting the alternative parameterization for cation species \cite{Dupont2013} did not lead to significant changes in the final energy and charge-voltage curves as shown in the Supporting Information. This is due to the fact that at a semiconductor surface, small changes in the surface charge lead to large potential shifts due to poor electrostatic screening, which implies that the voltage-dependent charge per atom remains small (on the order of a few tenths of the elementary charge), implying that the neutral-atom parameterization is more appropriate. The anion parameterization of Dupont \textit{et al.}\cite{Dupont2013} did lead to larger changes, but the qualitative trend remained the same. In the same vein, another parameterization has been developed by H{\"o}rmann \textit{et al.} \cite{Hormann2018} and tested by Nattino \textit{et al.}, demonstrating improved agreement between the calculated differential capacitance and experimental data \cite{Nattino2018}. As shown in the Supporting Information, using the SCCS parameterization of H{\"o}rmann \textit{et al.} does change the absolute voltages, but does not alter the voltage differences and thermodynamics trends significantly. 

Using the selected parameterization, we first simulate an electrically neutral system  with different dielectric responses on each side of the slab. These dielectric response are described by the experimental permittivities of silicon ($\epsilon_{\rm s} = 11.7$) and water ($\epsilon_{\rm m} = 78.3$). By aligning the potential inside the solution to the electrostatic reference of the continuum solvent, the flatband potential $\varphi_{\rm FB}$ can be directly related to the Fermi level $E_{\rm F}^{\circ}$ of the neutral surface through
\begin{equation}
	\varphi_{\rm FB} = -\frac{E_{\rm F}^{\circ}}{e_0}.
\end{equation}
where $e_0$ denotes the elementary charge. 

The charged electrode is then simulated by placing planar countercharges on each side of the slab and by assigning to the electrode a total charge $q$, split between the quantum-mechanical region $q^{\rm surf}$ and the bulk semiconductor $q^{\rm bulk}$ such that $q=q^{\rm bulk}+q^{\rm surf}$. (Here, the charge $q^{\rm bulk}$ is placed on the plane of the semiconductor side, while the charge $-q$ is assigned to the Helmholtz plane in the solution.) 

Finally, the long-range polarization of the depletion region is included by setting a frontier between the explicit and implicit semiconductor regions at a cutoff coordinate $z = z_{\rm c}$ along the transverse $z$-axis, typically two layers within the semiconductor electrode, corresponding to the inflection point of the potential profile where the net charge density vanishes (cf.~Poisson's equation). On the side of the cutoff bordering the solution, the electrode surface is described quantum mechanically, whereas on the other side, the Mott-Schottky model is used to describe the screening of the electrostatic field that arises from the dopant distribution (here, chosen to be $n$-type):
\begin{equation}
	n(\boldsymbol{r}) = N \exp\left[(\varphi_0-\varphi(\boldsymbol{r}))/(k_{\rm B}T)\right],
\end{equation}
where $N$ is the concentration of electron-donating defects, $k_{\rm B}$ is the Boltzmann constant,  $T$ is the ambient temperature, and $\varphi_0$ is the asymptotic value of the potential in the bulk semiconductor. (A similar equation with opposite charge applies for $p$-doped semiconductors.) It is important to note that this Mott-Schottky model circumvents the need to reparameterize the SCCS model on the semiconductor side, since the embedding contribution from the continuum semiconductor only serves as a intermediate step that is ultimately replaced by the Mott-Schottky medium. 

Using Poisson's equation with a Boltzmann distribution of electronic charges in the depletion layer, we can then derive the Mott-Schottky asymptotic relation of the electrostatic potential in the continuum region \cite{Gelderman2009}:
\begin{equation}
\left(\frac{\partial \varphi}{\partial z}\right)^2 = \frac{2 N}{\epsilon_0 } \left[ \varphi -  \varphi_0 + k_{\rm B}T \left( \exp\left(\tfrac{\varphi_0-\varphi}{k_{\rm B}T}\right) -1 \right) \right]
\label{eq:Mott-Schottky-1}.
\end{equation}
Then, noting that the exponential term is typically small (i.e., $\varphi-\varphi_0 \gg k_{\rm B}T$), the asymptotic value $\bar \varphi_0$ of the the difference between the electrostatic potential of the charged and neutral slabs $\bar \varphi$ can be obtained from
\begin{equation}
	\bar \varphi_0 = \bar {\varphi}(z_{\rm c}) -  k_{\rm B}T - \frac{\epsilon_0 }{2 N} \left(\frac{d \bar \varphi}{dz}(z_{\rm c})\right)^2.
	\label{eq:asymptote}
\end{equation}

\begin{figure}
	\includegraphics[width=0.5\columnwidth]{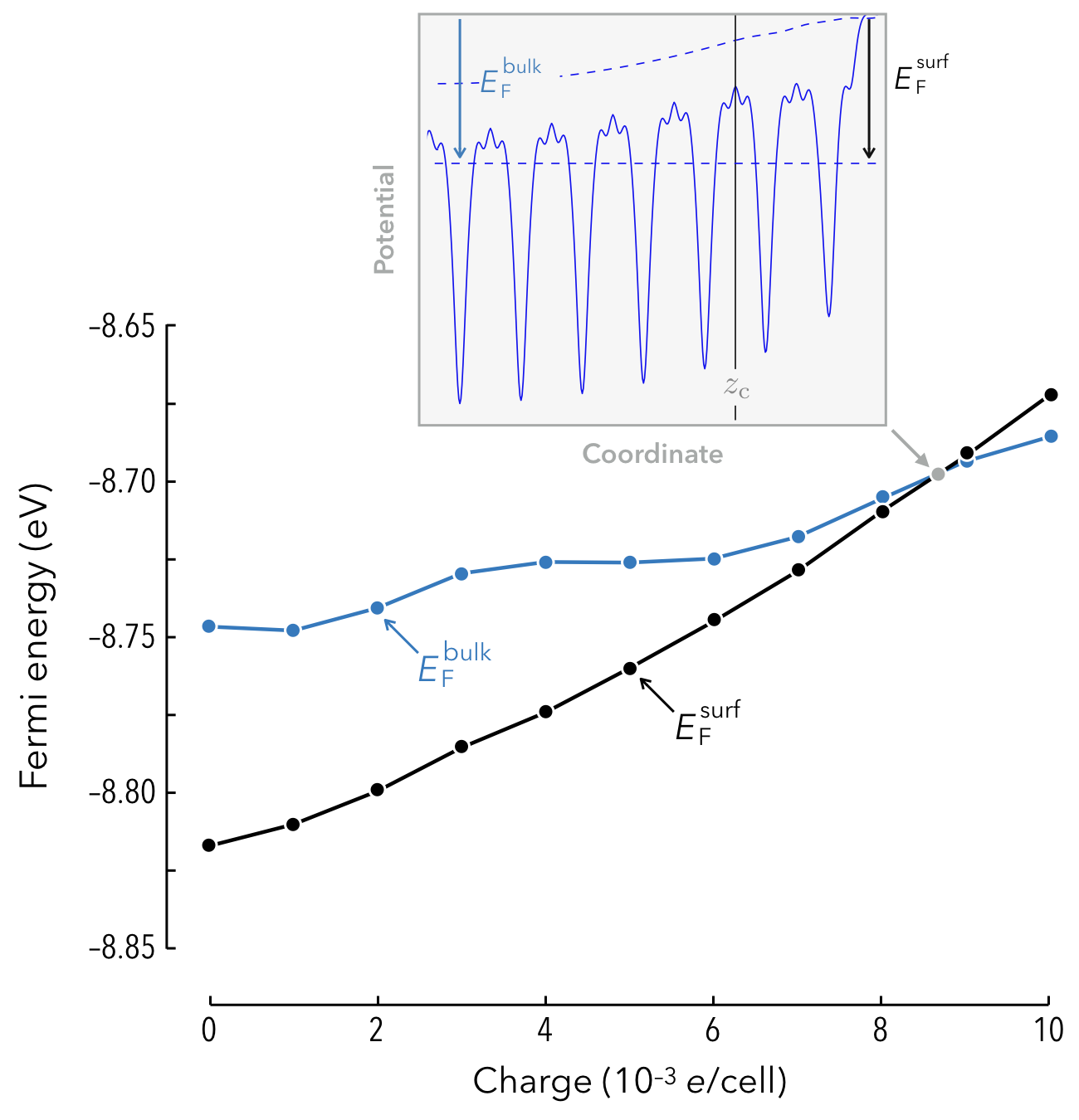}
	\caption{The equilibrium state of the silicon-water interface is found by computing the Fermi level of the bulk semiconductor $E_{\rm F}^{\rm bulk}$ and that of the surface region $E_{\rm F}^{\rm surf}$ as a function of the surface charge. At the intersections of these two electronic levels, the chemical potential of the electrons is constant across the entire system, as shown by the horizontal dashed line on the electrostatic profile of the inset.  Repeating this procedure for multiple net charges yields the charge-voltage response of the silicon electrode.
		\label{fig:fermi-levels}}
\end{figure}

As shown in Fig.~\ref{fig:fermi-levels}, the last step of the calculation consists of finding the charge distribution that self-consistently aligns the Fermi level of the explicit electrode surface ($E_{\rm F}^{\rm surf}$) to that of the bulk material ($E_{\rm F}^{\rm bulk}$). The latter is calculated by adding the asymptotic potential difference $\bar \varphi_0$ [Eq.~\eqref{eq:asymptote}] to Fermi level of the neutral slab:
\begin{equation}
	E_{\rm F}^{\rm bulk} = e_0 \bar \varphi_0 + E_{\rm F}^{\circ}.
\end{equation}
Once the equilibrium Fermi level is known, the voltage $\varphi = -E_{\rm F}/e_0$ and charge $q = q^{\rm bulk} + q^{\rm surf}$ can be calculated, allowing us to derive the equilibrium Schottky barrier height as a function of the applied voltage, as discussed in Section \ref{sec:sc-sol}. By dividing the total charge of the electrode $q$ by the surface area of the slab, we determine the charge density of each electrode surface.

\subsection{Computational details}

Electronic-structure calculations are performed using the {\sc quantum-espresso} software \cite{Giannozzi2009}. FA slab of five layer, with three of these layers geometrically constrained on the semiconductor side, is found to be sufficient to converge the Fermi level of the semiconductor-solution system within 50 meV. We center the slab in the supercell with a separation of 14 \AA\ between periodic slabs and optimize the geometry of the structure until interatomic forces are lower than 50 meV/\AA .  We use pseudopotentials generated with Perdew-Burke-Ernzerhof exchange correlation \cite{Perdew1996} with the projector augmented wave (PAW) method  from the {\sc sssp} library \cite{Lejaeghere2016}. The kinetic and charge density cutoffs are of 50 Ry and 750 Ry, respectively. The Brillioun zone is sampled with a shifted 5 $\times$ 5 $\times$ 1 Monkhorst-Pack grid and 0.03 Ry of Marzari-Vanderbilt smearing \cite{Marzari1997}.  We exploit the {\sc environ} module for the continuum solvent \cite{Andreussi2012}.

\section{Results}

\label{sec:sc-sol}
\begin{figure*}	
	\includegraphics[width=\textwidth]{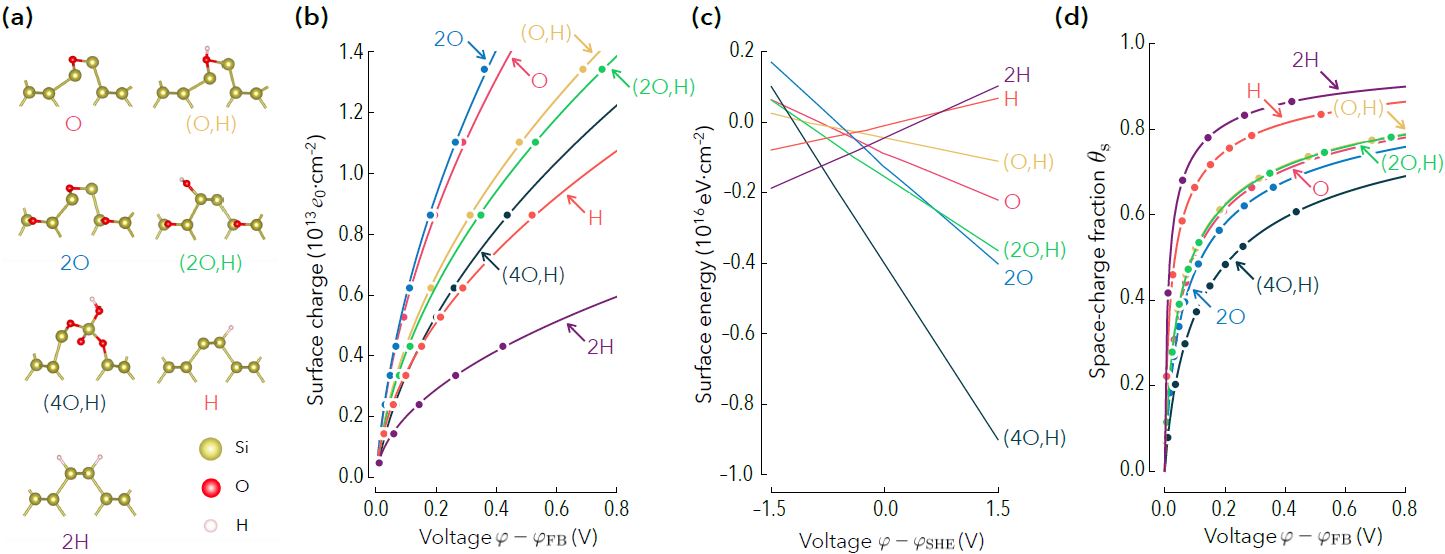}
	\caption{\small (a) Surface terminations of the silicon (110) electrodes, showing the configuration of the hydrogen and oxygen adsorbates. (b) Charge-voltage response of Si (110) structures with these adsorbates. Data points closely follow the proposed analytic model. (c) Surface free energy $\gamma$ of each silicon termination at pH = 7. The termination with the lowest free energy is the most stable at that potential. (d) Space-charge fraction $\theta$ of each electrode, the fraction of applied bias that develop in the depletion region. }
	\label{fig:chgvolt}
\end{figure*}

To understand the influence of the applied potential on the silicon-water interface, we simulate seven representative surfaces terminated by different combinations of oxygen and hydrogen species, namely, O, (O,H), 2O, (2O,H), (4O,H), H, and 2H, as shown in Fig.~\ref{fig:chgvolt}(a), with a semiconductor carrier concentration of 10$^{18}$ cm$^{-3}$ for the semiclassical Mott-Schottky model. The calculated charge-voltage responses are shown in Fig.~\ref{fig:chgvolt}(b). Here, the potentials are measured with respect to the flatband potential $\varphi_{\rm FB}$. These charge-voltage characteristics reveal that substantial differences arise from changing the surface termination due to charge pinning on the surface states associated with the dangling bonds that present at the interface. 

\begin{figure}
	\includegraphics[width=0.7\columnwidth]{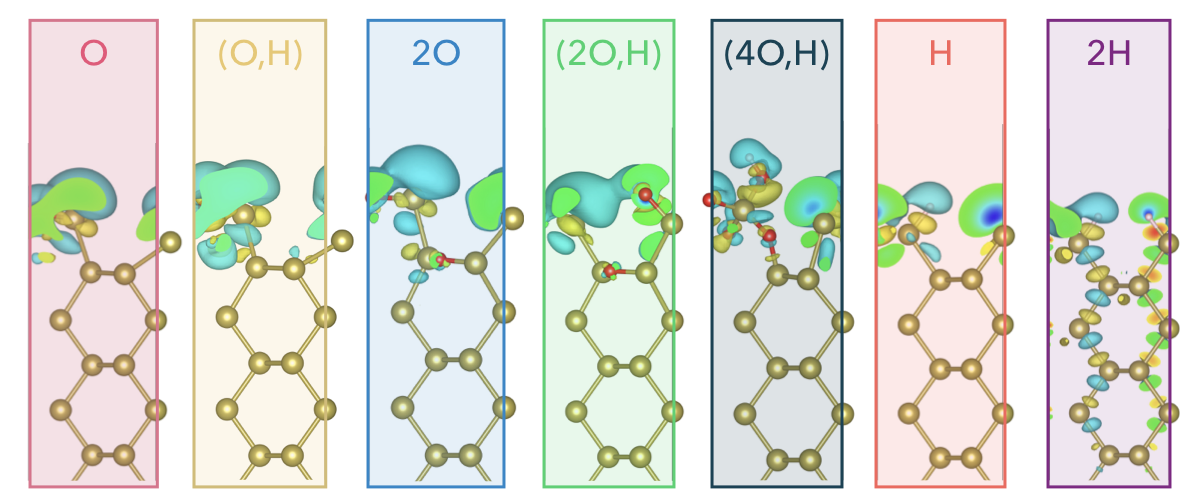}
	\caption{\small Accumulation of electronic charge in the explicit quantum mechanical portion of the electrode under bias. The isocontours indicate the positions of 90\% of the charge. When the majority of charge is at the surface, a significant surface dipole is formed, corresponding to a lower Schottky barrier. When charge is distributed across several layers (e.g., 2H), a smaller surface dipole is formed and a larger potential drop takes place across the space-charge region, inducing a higher Schottky barrier.}
	\label{fig:chg-acc}
\end{figure}

This observation is consistent with the fact that the charge of the depletion region causes a potential drop orders of magnitude larger than the surface dipole, due to poor electrostatic screening in the doped semiconductor. In contrast, surface terminations with significant charge accumulation (e.g, for the O and 2O covered surfaces), have most of their charge pinned at the interface. Figure \ref{fig:chg-acc} confirms this interpretation by showing the redistribution of the surface charge under applied voltage; for most of the surfaces, charge accumulates almost entirely on the adsorbate. For the 2H-terminated surface, however, the charge is distributed through all five layers of the simulated semiconductor, indicating that the charge extends deep into its depletion region. 

We then interpret the electrical response of the semiconductor-solution interface analytically by developing a model for predicting the charge-voltage relationship. This analytic model is premised on the fact that surface states contribute a metallic capacitance, as evidenced by examining the interfacial density of states. This surface-state capacitive contribution is then connected to that the ideal Mott-Schottky depletion layer, yielding
\begin{equation}
Q = \sqrt{2\epsilon_{0}\epsilon_{\rm s}e_{0} N \theta\varphi} + \mathscr C (1 - \theta)\varphi,
\label{eq:total-chg}
\end{equation}
where the first term is the contribution from the bulk semiconductor to the total charge and the second term is that of the surface states. Here, $\mathscr C$ is the capacitance of the surface states (which will be obtained by fitting the calculated charge-voltage response), $\epsilon_{s}$ is the dielectric constant of the semiconductor, $\varphi$ is the total potential drop across the electrode. We further introduce the space charge fraction, and $\theta = \varphi_{\rm s}/\varphi$, to quantify the extent to which surface states dominate the electrical response of the photoelectrode. This ratio represents the fraction of potential on the electrode that falls within the bulk of the semiconductor; this fraction is not a constant and varies as a function of the total potential across the electrode. 

An analytic expression for $\theta$ can be derived by noting that the amount of charge in the bulk semiconductor $q^{\rm bulk}$ is typically a constant fraction of the surface charge $q^{\rm surf}$ for all the terminations that we have studied (see Supporting Information). We thus introduce the coefficient $\eta = q^{\rm bulk}/q^{\rm surf}$ that describes the fraction of the charge that resides in the semiconductor section, allowing us to derive the following expression for $\theta$:
\begin{equation}
\theta = 1 + \phi/{\varphi} - \sqrt{ \phi^2/{\varphi}^2 + {2\phi/{\varphi}}},
\end{equation}
where $\phi = (\epsilon_{\rm s}\epsilon_0 e_0 \mathscr N)/(\eta^2 \mathscr C^2)$ is in units of volts and can be thought of as the switching potential, representing the point at which $\sim$25\% of the total potential drop takes place across the bulk semiconductor. At potentials above the switching potential, the fraction shifts rapidly such that the potential drop in the depletion layer makes up the majority of the total electrostatic drop. The charge-voltage model for the semiconductor-solution interface can now be fully described analytically by fitting the two constants $\eta$ and $\mathscr C$. Using this fitting procedure, Eq.~\eqref{eq:total-chg} can serve as an accurate analytic model of the charge-voltage response. 

It should be noted that this model is only valid at potentials less than the band gap of the material; if the potential exceeds this band gap, Zener tunneling will cause the semiconductor to act as a metal. The fitted values of $\eta$ and $\mathscr C$ for each adsorbate are reported in Table \ref{Table_1}, showing that the surface charge is more delocalized into the depletion region for the 2H termination ($\eta=17.3\%$) than it is for the O termination ($\eta =4.4\%$), causing the capacitance $\mathscr C$ to be higher in the former case. 

\begin{table}
	\small
	\centering
	\caption{Fitted surface state properties for the seven surface configurations tested. The fraction of the charge on the surface states that is on the bulk of the semiconductor is represented by $\eta = q^{\rm bulk}/q^{\rm surf}$. The capacitance of the surface state, assuming a metal like distribution, is represented as $\mathscr C$.}
	\begin{tabular*}{\columnwidth}{l @{\extracolsep{\fill}} cc}		     \\
	    \hline \hline \\
	    &  Charge ratio $\eta$ & Capacitance ${\mathscr C}$ ($\mu$F/cm$^{2}$) \\ \\
		\hline \\
		O  & 0.044 & 17.3 \\
		(O,H) & 0.060 & 14.0\\
		2O & 0.040 &  17.6 \\
		(2O,H) & 0.063 & 13.1\\
		(4O,H) & 0.067 & 8.1 \\
		H  & 0.087 & 14.8\\
		2H  & 0.173 & 10.5\\	
		
		\\ \hline \hline
	\end{tabular*}%
	\label{Table_1}%
\end{table}

With the charge-voltage relation in hand, we can calculate the surface free energy of each termination as a function of the applied potential using the Lippmann electrocapillary equation $\gamma = \gamma_0 - \int^{\varphi}_{\varphi_{\rm FB}} \sigma(\Phi)d\Phi  $.  Here, $\sigma$ is the charge per surface area of the electrode, $\gamma$ is the surface free energy of the charged slab at a certain potential $\varphi$, and $\gamma_0$ is the surface free energy of the slab under neutral charge conditions at the flatband potential. We determine the surface free energy of a neutral surface following the computational SHE method\cite{Man2011a,Norskov2004,Rossmeisl2007} by subtracting the energy of a surface terminated with adsorbates from the energy of the same surface without adsorbates and further subtracting the energy required to pull out a given adsorbate from the surrounding solution (see Sec.~S2 of the Supporting Information). We then calculate the free energy curves for each surface, as shown in Fig.~\ref{fig:chgvolt}(c). The structure with the lowest free energy at a given electrode potential is the thermodynamically stable configuration. Under most potential and pH conditions within the stability window of water, the (4O,H) configuration (which is the most oxidized termination tested) is the most stable, in agreement with the known tendency of silicon to oxidize in contact with water \cite{Runyan2013,Graf1989}.

Next, we calculate the Schottky barrier for each termination by determining the potential drop across the bulk semiconductor within our simulation, determined as $\varphi_{\rm s} = \bar \varphi_0 - \bar {\varphi}(z_{\rm c}) $, where $\bar \varphi_0$ is determined by Eq.~\ref{eq:asymptote} and $\bar {\varphi}(z_{\rm c})$ is determined from our calculations. As shown in Fig.~\ref{fig:chgvolt}(d), at low bias, the majority of the potential drop across the electrode is associated to the surface dipole $\delta\chi_{\rm s}^{\rm m}$ instead of the Schottky barrier, leading to a low space-charge fraction $\theta$. The fraction of the potential drop across the semiconductor increases rapidly with applied bias, then shows a much slower increase after $\sim$0.3 V. The surface termination with the least surface charge pinning (2H) unsurprisingly exhibits the largest fraction of the potential drop inside the semiconductor in the calculated voltage range.

\begin{figure}
	\includegraphics[width=0.5\columnwidth]{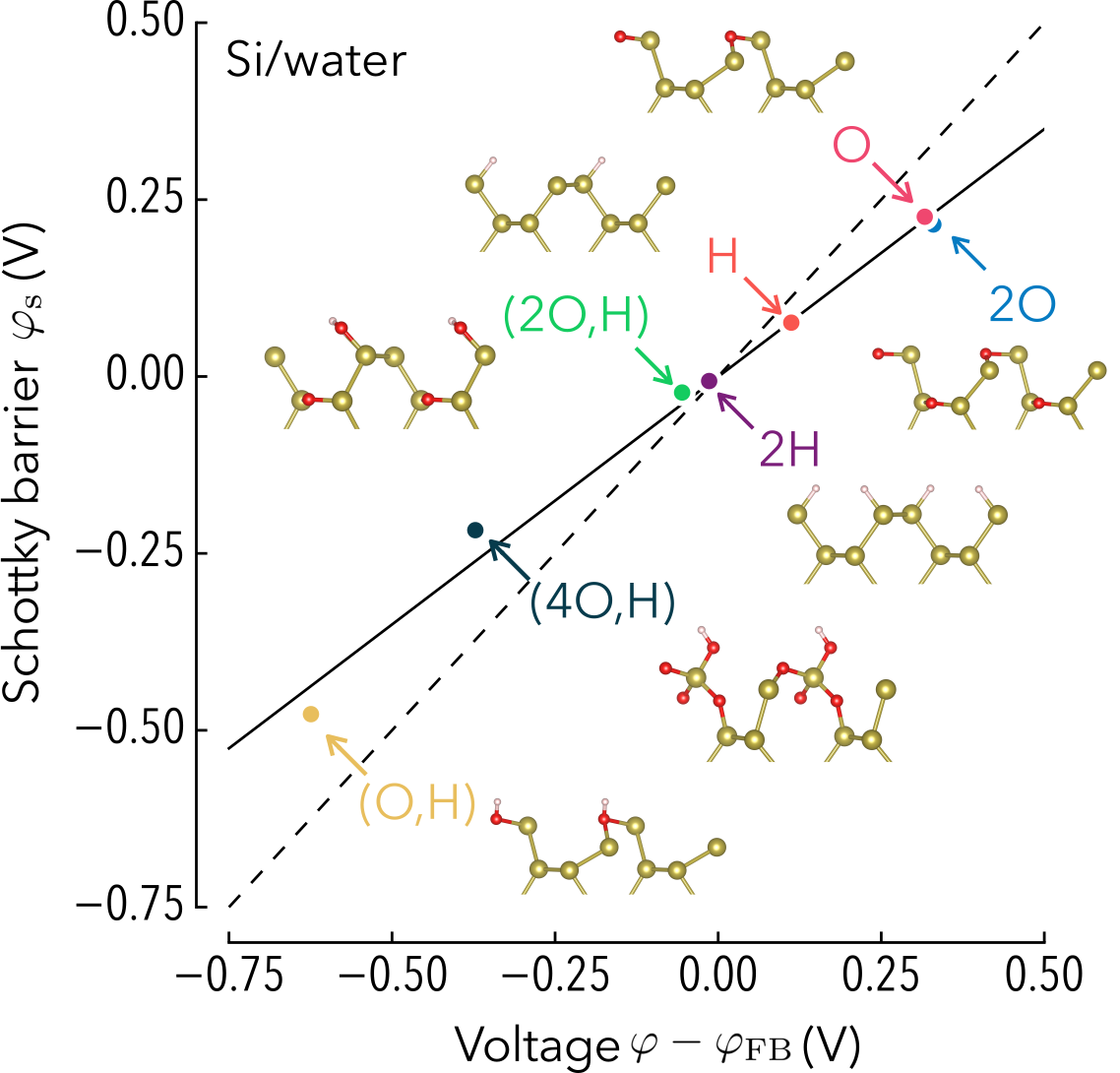}
	\caption{ The equilibrium barrier height as a function of the difference between the hydrogen evolution potential and the flatband potential. An ideal semiconductor junction, would show a unit slope, $\mathscr{S}=1$ (dashed line). Due to charge trapping by surface states, the Schottky barrier heights are significantly reduced, corresponding to a charge-pinning fraction equal to $\mathscr{S} = 0.7 \pm 0.005$ for the silicon-water system.}
	\label{fig:Schottky}
\end{figure}

Having computed the Schottky barrier as a function of external bias, we can now calculate an important parameter, the charge-pinning fraction $\mathscr{S}$,  which describes the reduction of the Schottky barrier from its theoretical maximum due to charge pinning  \cite{Kurtin1969a,Brillson1978,Tung2014,Natarajan1998,Bard1980,Cowly1965}. Explicitly, the charge-pinning fraction is defined (for n-type semiconductors) as
\begin{equation}
\varphi_{\rm s} = \mathscr{S} (\chi_{\rm m} - \chi_{\rm s}),
\end{equation}
where  $\mathscr{S}=1$ corresponds to an ideal semiconductor junction with no charge trapping, while a value of $\mathscr{S}=0$ indicate that all the charge accumulates in surface states, inducing complete Fermi-level pinning. Here, $\chi_{\rm m}$ and $\chi_{\rm s}$ are the vacuum referenced electronegativity of the embedding medium and the semiconductor, respectively, which can be calculated as the opposite of the Fermi level when the vacuum section of the slab is aligned to zero \cite{Weitzner2017,Keilbart2017}. The value of $\mathscr{S}$ is typically obtained experimentally by measuring the Schottky barrier of a semiconductor against different chemical environments and calculating the slope of the resulting line of best fit. 

To calculate the equilibrium Schottky barrier $\varphi_{\rm s}$ and charge-pinning factor $\mathscr{S}$, we first find the applied bias that sets the voltage of the electrodes to the hydrogen evolution potential. We then directly extract the Schottky barrier height at this applied bias by calculating the potential drop within the bulk semiconductor. By plotting the Schottky barrier height as a function of the applied bias to bring the interface into equilibrium with the hydrogen evolution redox level, we can find the charge-pinning fraction from the slope, as shown in Fig.~\ref{fig:Schottky}. By linear regression, we measure a slope of $\mathscr{S} \approx 0.7$, which corresponds to a significant deviation from the ideal trend of $\mathscr{S}=1$, reflecting the contribution from the surface dipole to the renormalization of the Schottky barrier height.

Our work highlights some of the contradictory requirements that limit the photocatalytic activity of silicon photoelectrodes. For the hydrogen evolution reaction, the ideal Schottky barrier would be positive --- driving electrons from the bulk of the electrode to the surface. However, the two terminations with the most positive Schottky barriers (one oxygen and two oxygens adsorbed onto the surface) are both unstable in the redox window of water as shown in Fig.~\ref{fig:chgvolt}c. The difficulty of simultaneously achieving surface stability and effective charge transfer across the silicon-water interface provides quantitative insights into the limited photocatalytic activity of silicon photoelectrodes in an aqueous environment. It is interesting to note that this interpretation requires only a single layer of oxide to form on the silicon surface and is not predicated on the formation of a thick deposit of silica. Instead, the low photocatalytic performance of silicon is here possibly explained by the surface states induced by an atomically thin oxide layer, suggesting a much more drastic influence on the solar-to-hydrogen conversion efficiency. 

\section{Conclusion}

We have examined the performance of silicon for water splitting as a function of the exposed termination of the semiconductor electrode from first principles. We have developed a methodology for predicting Schottky barriers accurately, taking into account the interactions between the charge pinned in the interfacial region and the charge accumulated within the depletion layer of the semiconductor. We have applied this methodology to predict the  stability, Schottky barrier, and charge-pinning fraction of different surface terminations for silicon in water. Our study shows that the structures with the most favorable Schottky barriers for water splitting are electrochemically unstable, shedding light on the physical origins of their low solar-to-hydrogen conversion efficiency. Our work demonstrates the broad capabilities of the SCCS model, which can be used in conjunction with recent developments in predicting accurate electronic structures \cite{Nguyen2017,Timrov2018}, to predict and understand the photocatalytic activity of passivated semiconductor electrodes from first principles.

\section{Supplementary Information}
The Supplementary Information section provides a discussion of electrostatic potential references, further details on the model, and further explanations on the calculation of surface stability.

\begin{acknowledgements}
	The authors acknowledge primary support from the National Science Foundation under Grant DMR-1654625, and partial support from the 3M Graduate Fellowship and Penn State University Graduate Fellowship. The authors gratefully acknowledge Giulia Galli, H\'ector D.  Abru\~na, Roman Engel-Herbert, and Suzanne Mohney for fruitful discussions.
\end{acknowledgements}

\end{document}


\title{Supporting Information: ``Electrochemical stability and light-harvesting ability of silicon photoelectrodes in an aqueous environment''}
	
\author{Quinn Campbell}
\author{Ismaila Dabo}
\affiliation{Department of Materials Science and Engineering, Materials Research Institute, and Penn State Institutes of Energy and the Environment, The Pennsylvania State University, University Park, PA 16802, USA}

\maketitle
\section{Electrostatic references}
\begin{figure*}	
	\includegraphics[width=\textwidth]{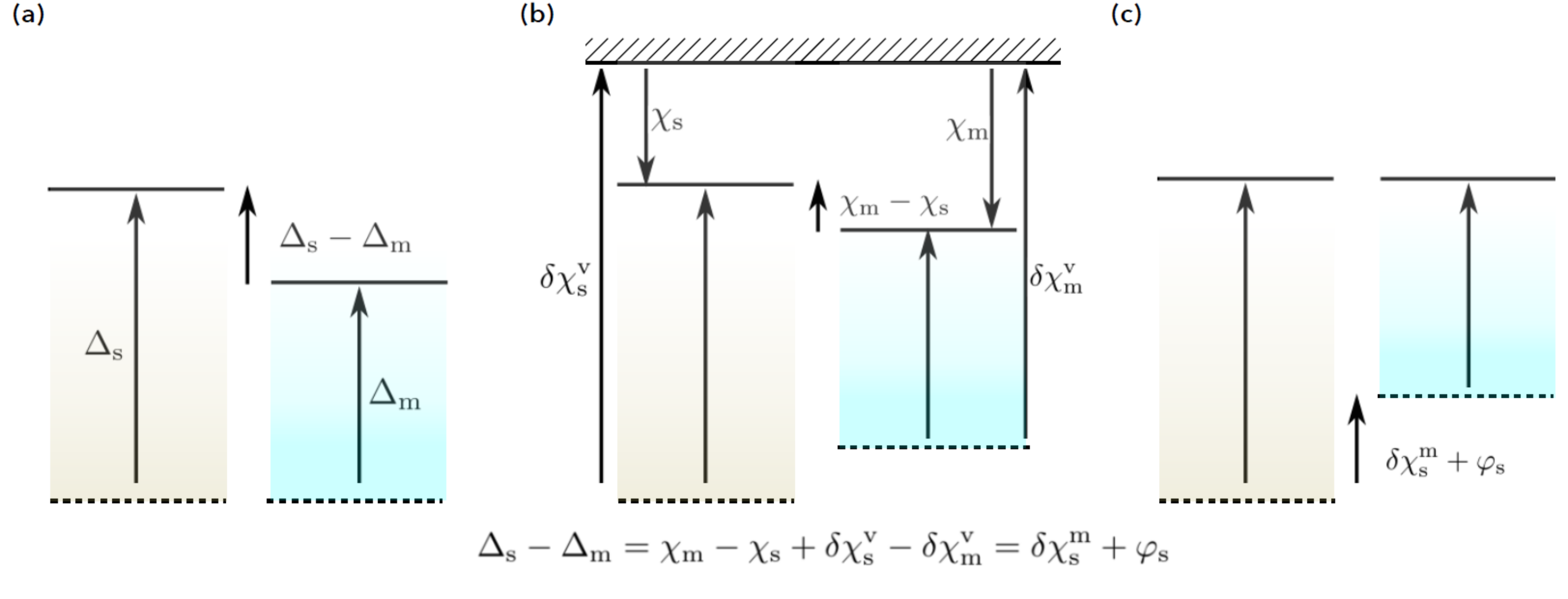}
	\caption{\small Three possible conventions in describing a semiconductor junction. (a) Alignment of the semiconductor and medium by the average bulk electrostatic potential. The difference between the Fermi levels in this alignment can be characterized as $\Delta_{\rm s} - \Delta_{\rm m}$ where $\Delta_{\rm s}$ and $\Delta_{\rm m}$ are the difference between the Fermi level and the bulk electrostatic levels of the materials. (b) Alignment by the vacuum level. Here, the primary values are the electronegativity of the medium $\chi_{\rm m}$ the semiconductor $\chi_{\rm s}$. (c) Alignment by Fermi levels, as would take place in an equilibrium system. The difference between the two bulk electrostatic levels can now be quantified as the surface dipole $\delta \chi_{\rm s}^{\rm m}$ in addition to the Schottky barrier $\varphi_{\rm s}$. With careful alignment, all these methods can be used for calculating Schottky barriers.}
	\label{fig:references}
\end{figure*}

Figure \ref{fig:references} shows three different conventions for aligning potentials between the semiconductor and the embedding medium. While in the context of first-principles calculations, potentials have often been aligned based on the bulk electrostatic levels as shown in Fig.~\ref{fig:references}a, these potentials are typically defined experimentally in terms of the electronegativity of the medium $\chi_{\rm m}$ and the semiconductor $\chi_{\rm s}$, where electronegativity is defined as the energy of moving one electron from the material to vacuum, as shown in panel b. In reality, of course, after equilibration, it is the Fermi levels that match one another as shown in panel c. The difference between the two bulk electrostatic levels can now be quantified as the surface dipole $\delta \chi_{\rm s}^{\rm m}$ in addition to the Schottky barrier $\varphi_{\rm s}$. Here, the surface dipole can be expanded $\delta \chi_{\rm s}^{\rm m} = \delta\chi_{\rm s}^{\rm v}- \delta\chi_{\rm m}^{\rm v} + \Delta V $, splitting the interfacial dipole into a component that is the difference between the semiconductor and medium's bulk electrostatic levels and a component $\Delta V$ due to rearrangement of atoms and any charge accumulation and/or transfer at the surface. All of these methods can be used to find the Schottky barrier as long as one is careful with alignment, specifically noting that the difference between electronegativities does not take into account the voltage difference between the bulk electrostatic levels of the semiconductor and embedding medium, i.e., $\delta\chi_{\rm s}^{\rm v} - \delta\chi_{\rm m}^{\rm v} $.

\section{Role of the interfacial charge}
\label{sec:models}
\begin{figure}
	\includegraphics[width=0.5\columnwidth]{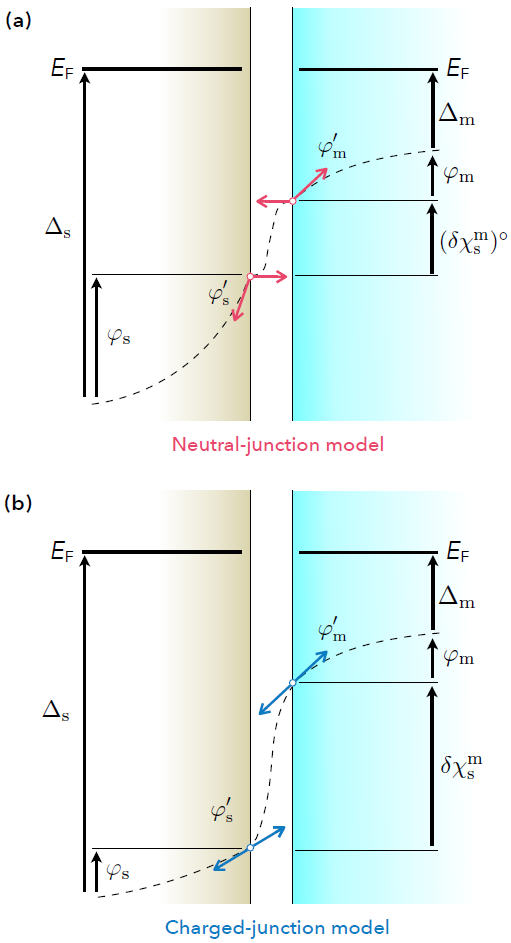}
	\caption{(a) The neutral-junction model for the calculation of Schottky barriers. Here, the neutral-junction simulated has a flat potential at the edges, leading to a discontinuity in the potential slope between the calculated junction and the assumed Mott--Schottky potential distribution. In contrast, the charged-junction model (panel b) simulates charge within the explicit interface, leading to a continuous derivative of the macroscopic potential and a modified predicted surface dipole and Schottky barrier.
		\label{fig:models}}
\end{figure}
In the neutral-junction model for calculating Schottky barriers\cite{Ping2015,Farmanbar2015,Goniakowskt2004,Stengel2011a,Khomyakov2009,Dong2008,McKee2003}, the Fermi levels of the semiconductor and the medium are aligned. From there, a potential difference can arise between the average electrostatic potential of the bulk medium $\Delta_{\rm m}$ and the electrostatic potential of the bulk semiconductor $\Delta_{\rm s}$. A neutral, unbiased junction is simulated using electronic-structure methods and the interfacial dipole $(\delta \chi_{\rm s}^{\rm m})^\circ$ is calculated. The remainder of the difference between $\Delta_{\rm s}$ and $\Delta_{\rm m}$ is then equated to the Schottky barrier $\varphi_{\rm s}$. However, since the neutral junction has no net charge, the slope of the potential at each end of the junction can be determined to be zero (Gauss' law). This leads to a mismatch of the electrostatic slope between the neutral junction and the depletion region of the semiconductor (and a similar mismatch between the potential slope on the medium side, depending on the specific distribution of charges within the medium) as depicted in Fig.~\ref{fig:models}a.

For a self-consistent solution to the Schottky barrier problem, the slope of the potential at the edges of the interface must match the slope of the potential at the Mott--Schottky and medium electrostatic potential distribution junction. This implies that charge will accumulate at the interface, making the simulation of a charged junction necessary. This change will lead to modifications in the surface dipole $\delta \chi_{\rm s}^{\rm m}$ and thus the Schottky barrier $\varphi_{\rm s}$, as depicted in Fig.~\ref{fig:models}b. For the charged-junction model to be effective, it is necessary to implement a procedure for determining the charge on the interface that will allow the slope of the potential to be continuous throughout the interface. The next section describes such a method.

\begin{figure}
	\includegraphics[width=0.5\columnwidth]{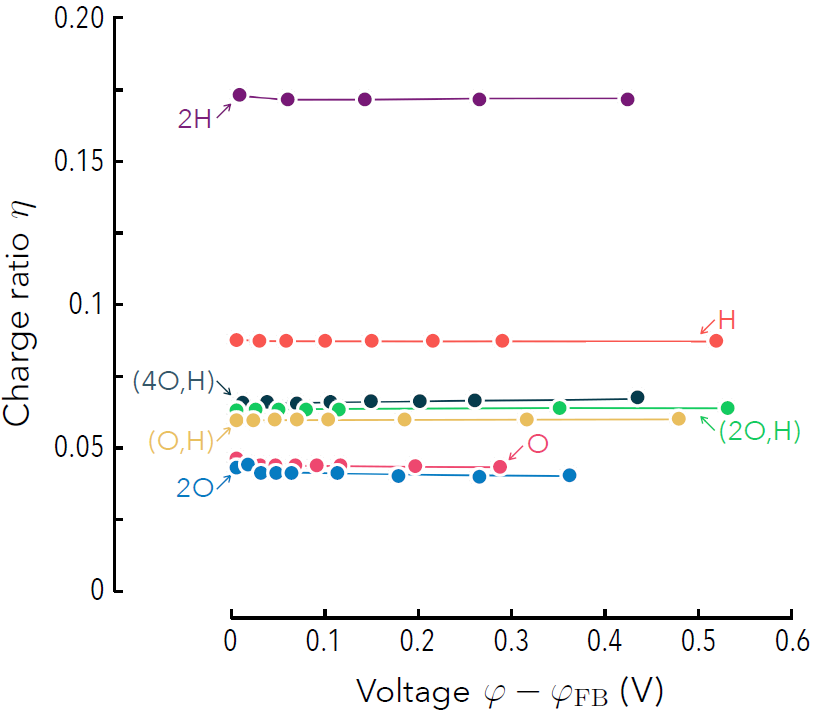}
	\caption{ The ratio $\eta = q^{\rm bulk}/q^{\rm surf}$ remains constant as a function of potential, enabling one to derive an analytical expression for the charge on a semiconductor electrode.}
	\label{fig:eta-constant}
\end{figure}

\section{Varying parameterizations of the self-consistent continuum solvation (SCCS) model }

Within this work, we utilize the self-consistent continuum solvation (SCCS) model with the parameterization suggested by Andreussi \textit{et al.}, where $\rho_{\rm max}= 5 \times 10^{-3}$ a.u., $\rho_{\rm min}= 1 \times 10^{-4}$ a.u., $\gamma = 72.0$ dyn/cm, $\alpha = -22$ dyn/cm, and $\beta = -0.35$ GPa \cite{Andreussi2012}. This parameterization was developed by fitting the solvation energy of neutral molecules. Alternate parameterizations have been proposed, notably for the case of charged anion and cation molecules in water \cite{Dupont2013}. For the case of cations in water, the parameterization of $\rho_{\rm max}= 0.0035 $ a.u., $\rho_{\rm min}= 2 \times 10^{-4}$ a.u., $\gamma = 25.0$ dyn/cm, $\alpha = -20$ dyn/cm, and $\beta = 0.125$ GPa is used, whereas for anions, the parameterization of $\rho_{\rm max}= 0.0155$ a.u., $\rho_{\rm min}= 0.0024$ a.u., $\gamma = 22.45$ dyn/cm, $\alpha = -22$ dyn/cm, and $\beta = 0.0$ GPa is employed. Recently H{\"o}rmann \textit{et al.} \cite{Hormann2018} have also developed a new parameterization based on the case of a Pt slab in water where $\rho_{\rm max}= 0.01025 $ a.u., $\rho_{\rm min}= 1.3 \times 10^{-3}$ a.u., and $\alpha$, $\beta$, and $\gamma$ are set to zero to avoid unphysical volume contributions to the energy of the system. 

\begin{figure}
	\includegraphics[width=0.5\columnwidth]{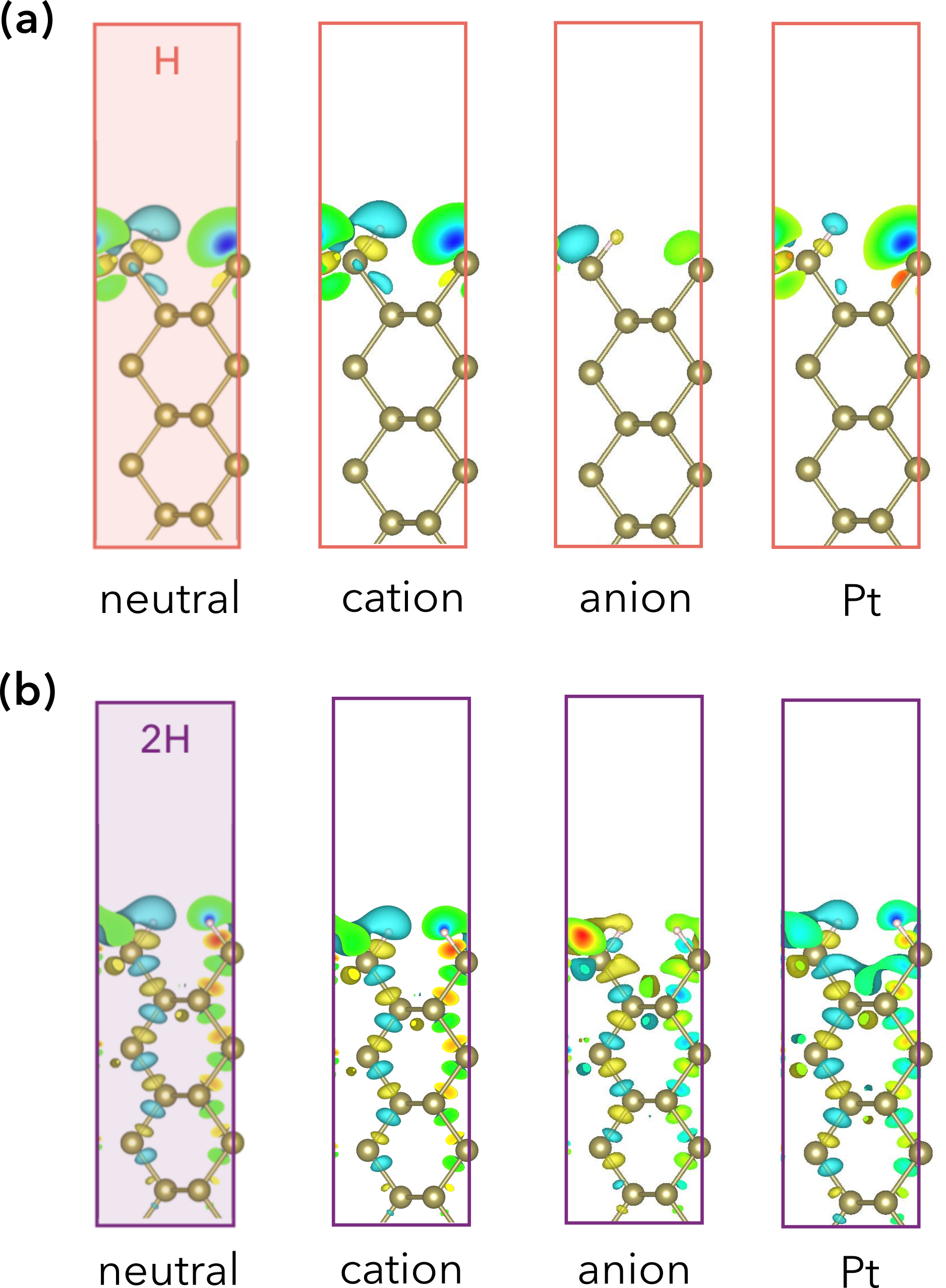}
	\caption{ The distribution of added charge found by varying the parameterization of the SCCS model for both (a) one hydrogen and (b) two hydrogens adsorbed onto the surface.}
	\label{fig:varying-chg-distr}
\end{figure}

To asses the impact varying the parameterization of the SCCS model has on the final result of our calculations, we tested the silicon (110) interface with both one and two hydrogens adsorbed with a total electrode charge of $q = 0.005$ $e_0$. The resulting charge distributions are shown in Fig.~\ref{fig:varying-chg-distr}. The cation parameterization leaves the charge on the electrode essentially unchanged. Since the amount of charge being added to the slab is so minimal for semiconductor electrodes, the effect of changing to a positive charged and uncharged parameterization is minimized. The anion parameterization, however, does lead to more significant change. To better match the conditions under which the anion parameterization was developed, we instead apply a charge of $q = -0.005$ $e_0$. This parameterization pushes the dielectric region closer in to the surface and results in more localized charge at the surface, leading to lower electrode potentials. The parameterization based on Pt demonstrates the largest difference between the neutral water parameterization, though large changes are still avoided. In general, slightly more charge accumulates at the surface using the Pt model, though the trend of charge trapping between different adsorbates remains the same. This physical intuition can be confirmed by numerically examining the resulting potentials from each parameterization in Table \ref{Table_I}. Since more charge gets trapped at the surface when using the Pt parameterization of the SCCS model, less charge is distributed throughout the bulk semiconductor, leading to lower voltages for the electrode. Importantly, however, the qualitative trend between different adsorbates remains the same for all parameterizations.

\begin{table}
	\small
	\centering
	\caption{The bulk potential of an electrode with a total charge of 0.005 $e_0$ added per unit cell, with varying parameterizations of the SCCS model.}
	\begin{tabular*}{\columnwidth}{l @{\extracolsep{\fill}} ccc}			     \\
	    \hline \hline \\
	    Surface Adsorbate & charge ($e_0$) &  SCCS parameterization & Voltage $\varphi - \varphi_{\rm FB}$ (V) \\ 
		\hline \\
		H & 0.005 & neutral & 	0.0576 \\
		  & 0.005 & cation  &   0.0593 \\
		  & -0.005 & anion   &   -0.0299   \\
		  & 0.005 & Pt      &   0.0136 \\
	   2H & 0.005 & neutral & 	0.1435 \\
		  & 0.005 & cation  &   0.1472 \\
		  & -0.005 & anion   &   -0.0445  \\
		  & 0.005 & Pt      &   0.0712 
		
		\\ \hline \hline
	\end{tabular*}%
	\label{Table_I}%
\end{table}

\section{Derivation of surface energy}

The free surface energy of a surface with $N_{H}$ hydrogen adsorbates, $N_{O}$ oxygen adsorbates, and $q$ surface free charges can be found as follows: 
\begin{equation}
G(N_{\rm H},N_{\rm O},q) =  G(N_{\rm H},N_{\rm O},0) + \int_{0}^{q} \varphi(q') dq', 
\label{eq:surface-energy-vs-charge}
\end{equation}
where $\Phi$ represents the electrical potential of the interface with $q'$ free charges on it. To find $G(N_H,N_O,0)$ we take the DFT energy of this structure, $E(N_H,N_O,0)$, and subtract the DFT energy of a silicon surface with no adsorbates, $E(0,0,0)$. We additionally subtract the energy of taking a hydrogen ion $\mu(H^+)$ or hydroxyl ion $E(OH^{-})$ out of solution:
\begin{equation}
	G(N_{\rm H},N_{\rm O},0) = E(N_{\rm H},N_{\rm O},0) - E(0,0,0) - (N_{\rm H}-N_{\rm O})  \mu(H^+) - N_{\rm O} \mu(OH^{-}).
\end{equation}
The energy of hydrogen ions in solution can be found from the following reaction 
\begin{equation}
	H^+_{aq} + e^- \leftrightarrow 1/2 H_2(g)
\end{equation}
which is at equilibrium at the potential of the reversible hydrogen electrode. Therefore the energy of $H^+$ is 
\begin{equation}
	\mu(H^+) = 1/2 E(H_2) - \varphi ({\rm SHE}) - 0.06 {\rm pH}.
\end{equation}
where $E(H_2)$ is the DFT energy of hydrogen gas.

Similarly the energy of $O^{2-}$ can be found by first finding the energy of $OH^-$ taken out of solution. The energy of $OH^-$ can be found from the reactions
\begin{equation}
 1/2 O_2 (g) + 2H^+ + 2e^- \leftrightarrow H_2O
\end{equation}
and
\begin{equation}
	1/2 O_2 + H_2O + 2e^- \leftrightarrow 2OH^-
\end{equation}
which are at equilibrium at the water splitting potential 1.23 V (RHE).

Combining these equations and plugging in the previously calculated formula for $H^+$ gives
\begin{equation}
	\mu(OH^-) = 1/2E(H_2) + 1/2E(O_2) + \varphi ({\rm SHE}) -2*1.23 V+ 0.06{\rm pH}.
\end{equation}
where $E(O_2)$ is the DFT energy of an oxygen molecule.

With these equations in hand, the free energy at the surface can now be easily calculated as a function of pH and voltage. We went on to calculate a pourbaix diagram of the most stable phase at several solution pH values and voltages as shown in Figure S3. It is important to note that we only found the lowest energy of the seven different adsorbate configurations tested, assuming the number of adsorbates stays the same and configurational entropy is not included. This model could be extended to take into account the full change in adsorbate coverage with potential with a Monte Carlo or cluster expansion model. 

\begin{figure}
	\includegraphics[width=0.5\columnwidth]{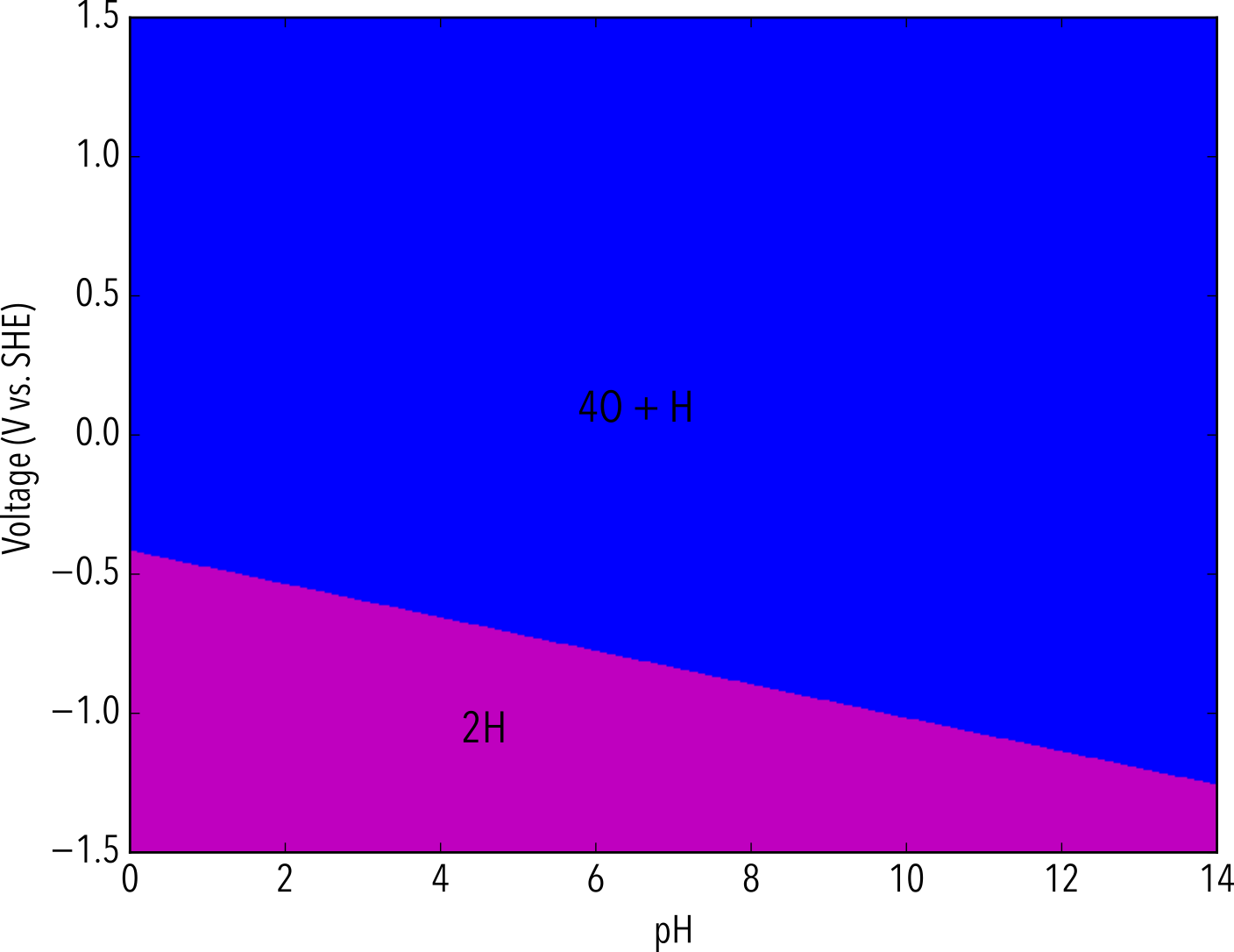}
	\caption{ The calculated Pourbaix diagram for the Si (110) surface in contact with water, accounting for the energy from charge trapping at the semiconductor--solution interface. This diagram shows what adsorbate structure will be most stable at different values of the electrode potential and solution pH. At low pH, Si will be most stable with two hydrogen adsorbed, but at the majority of normal conditions, the most oxidized form of silicon tested will be the most stable, matching experimental observations of the rapid oxidation of silicon in water.}
\end{figure}